\documentclass[superscriptaddress,twocolumn,prb,showpacs,preprintnumbers,amsmath,amssymb]{revtex4}

\usepackage{graphicx}
\usepackage{dcolumn}
\usepackage{bm}
\usepackage{ulem}

\usepackage[USenglish]{babel}
\usepackage{color}
\usepackage{verbatim}
\usepackage{amssymb}

\begin{document}

\title{ ~Electron-phonon interaction and spectral weight transfer in Fe$_{1-x}$Co$_{x}$Si
}

\author{D. Menzel}
\thanks{The first two authors contributed equally and share the first authorship of this work}
\affiliation{Institut f\"ur Physik der Kondensierten Materie, Technische Universit\"at Braunschweig, Mendelssohnstr. 3, D-38106 Braunschweig, Germany}
\author{P. Popovich}
\thanks{The first two authors contributed equally and share the first authorship of this work}
\affiliation{Max Planck Institute for Solid State Research, Heisenbergstr. 1, D-70569 Stuttgart, Germany}
\author{N. N. Kovaleva}
\affiliation{Max Planck Institute for Solid State Research, Heisenbergstr. 1, D-70569 Stuttgart, Germany}
\affiliation{Department of Physics, Loughborough University, Loughborough, LE11 3TU, United Kingdom}
\author{J. Schoenes}
\affiliation{Institut f\"ur Physik der Kondensierten Materie, Technische Universit\"at Braunschweig, Mendelssohnstr. 3, D-38106 Braunschweig, Germany}
\author{K. Doll}
\affiliation{Max Planck Institute for Solid State Research, Heisenbergstr. 1, D-70569 Stuttgart, Germany}
\author{A. V. Boris}
\affiliation{Max Planck Institute for Solid State Research, Heisenbergstr. 1, D-70569 Stuttgart, Germany}
\affiliation{Department of Physics, Loughborough University, Loughborough, LE11 3TU, United Kingdom}

\date{\today}

\begin{abstract}

A comprehensive ellipsometric study was performed on  Fe$_{1-x}$Co$_{x}$Si single crystals in the spectral range from 0.01~eV to 6.2~eV. Direct and indirect band gaps of 73 meV and 10 meV, respectively, were observed in FeSi at 7~K. One of four infrared-active phonons that is energetically close to the direct absorption edge is coupled both to the electrons and to the low-energy phonon. This is evident from asymmetry in the phonon line shape and a reduction of its frequency when the absorption edge shifts across the phonon energy due to the temperature dependence of the direct band gap. As the temperature increases, the indirect gap changes sign, which manifests as a transition from a semiconductor to a semimetal. The corresponding gain of the spectral weight at low energies was recovered within an energy range of several eV. The present findings strongly support the model indicating that  Fe$_{1-x}$Co$_{x}$Si can be well described in an itinerant picture, taking into account self-energy corrections.
\end{abstract}

\pacs{71.20.Be, 78.20.Ci, 78.30.Am}

\maketitle

\section{\label{sec:Intro}Introduction}

Due to its unusual properties, the narrow-gap semiconductor FeSi continues to be the subject of intriguing debate among theorists and experimentalists with regard to its electronic nature. Some groups have described this material as a correlated system with local Fe moments similar to those of the compounds known as Kondo insulators.\cite{aeppli92} FeSi has an insulating non-magnetic ground state, and its magnetic susceptibility increases with rising temperature. \cite{jaccarino67} This behavior has been modeled by two narrow peaks in the density of states (DOS) on both sides of the gap.\cite{mandrus95} However, local density approximation (LDA) calculations do not predict such peaks in the DOS, and this disagreement has been construed as an indication of correlation effects.\cite{mattheiss93} The fact that a low temperature mass enhancement of about 30 free electron mass has been found in specific heat measurements \cite{chernikov97} lends additional support to the argument for strong electronic correlations. Results from Raman scattering have been interpreted in terms of strong coupling between charge excitations and the lattice. \cite{nyhus95} 

More recent results on high-purity samples indicate that FeSi can be well described in an itinerant picture. Recent photoemission (PE) investigations of  FeSi single crystals have not revealed any indication of  the existence of a Kondo resonance close to the Fermi energy.\cite{zur07,klein08} Instead, good agreement has been found between photoemission experiments and band structure calculations using density functional theory;\cite{neef06} this agreement favors  an itinerant description rather than a Kondo model. Raman spectroscopy on the same FeSi single crystals reveals that the linewidth of an $E$-phonon increases with growing temperature \cite{menzel06,racu07} due to the shift of the chemical potential with increasing temperature. Also, magnetic measurements as a function of pressure on Co-substituted FeSi can be interpreted in an itinerant picture in which the magnetic behavior is strongly influenced by a small shift in the electronic density of states.\cite{menzel06} Doping FeSi with more than 5 at.\% Co results in a metallic state due to the closing of the band gap. Between 5 and 80 at.\% Co, ferromagnetic order is observed and the sample displays a conical spin structure.\cite{menzel04} Above 80 at.\%  Fe$_{1-x}$Co$_x$Si is paramagnetic, and finally, CoSi is a diamagnetic semimetal.\cite{wernick72}

So far, results from optical and thermodynamic studies have not been conclusive. The absolute value of the experimentally derived band gap varies between 0 and 100 meV; clearly, the correct size of the intrinsic band gap has still not been reliably determined. Two reasons can be proposed for this. First, the quality of the samples plays a crucial role in determining the electronic properties. Impurities lead to an only partially developed gap with residual states at the Fermi energy, as observed in earlier high-resolution photoemission experiments.\cite{breuer97} The samples used in the present study have a very low impurity concentration of 0.17 at.\% and display a fully developed gap in the density of states.\cite{zur07,klein08} Second, in past studies optical gaps have been determined using reflectivity measurements.\cite{chernikov97,schlesinger93,degiorgi94} However, accurate absolute values of the dielectric function are difficult to obtain with this method, especially when the reflectivity is close to unity in the far-infrared range. Moreover, for absolute values the comparison with a reference sample is always necessary and for a correct Kramers-Kronig analysis, values toward zero and infinite energy must be extrapolated. A much better technique is the direct measurement of both the real and the imaginary part of the dielectric function without the need for a reference sample, which was done in this work via spectroscopic ellipsometry. 

Reflectivity measurements showed that the band-gap disappears at a much lower temperature than expected. When FeSi is cooled down, the band gap is formed and, consequently, spectral weight at low energies is lost. Some experiments indicate that this spectral weight does not appear just above the gap, as expected in a classical band-like material, but is redistributed over a large energy scale, more than 80 times the gap energy.\cite{chernikov97,schlesinger93,damascelli97} This is also observed in, for example, CePd$_3$, which is considered a Kondo-insulator type material and where a similar shift in the spectral weight from low to high energies is found.\cite{webb86} This has led to the conclusion that FeSi is a strongly correlated insulator. However, based on other reflectivity experiments on FeSi single crystals, it has been claimed that the missing spectral weight below the gap energy is redistributed around the gap edge, which is a common property of a conventional semiconductor.\cite{degiorgi94} We believe that this disagreement is due to the difficulty in obtaining absolute values from reflectivity measurements.
 
  Another key issue is the determination of the ionicity and the charge transfer between the constituents, which can also be easily derived from the dielectric function. For these quantities, accurate determination of the absolute values from ellipsometric data is again mandatory.

 In this paper, we report on a decisive examination of the optical properties of Fe$_{1-x}$Co$_x$Si by spectroscopic ellipsometry from the far-infrared (20 meV) to the UV spectral range (6.2 eV). Section II describes the experimental details. In Section IIIA, eigenfrequencies and lineshapes of infrared active modes are determined and assigned. In Section IIIB, the determinations of direct and indirect band gaps as functions of Co concentration and temperature are presented. In Section IIIC, the electron-phonon coupling is considered as the origin of phonon anharmonicity. The self-consistency of the spectra is verified using the generalized Lyddane-Sachs-Teller relation. The Born and Szigeti effective charge is calculated to determine the charge transfer between Fe and Si. In Section IIID, we address the question of what causes the enhancement of the far-IR spectral weight due to the optical gap closing. We report that this additional spectral weight is transferred mostly from the high-frequency region of 0.5-2.5~eV. Finally, our conclusions are drawn in Section IV.

\section{Experimental details}
For the preparation of the Fe$_{1-x}$Co$_x$Si samples 99.98\,\% Fe, 99.9+\,\% Co, and Si ($\rho_n = 300\ \Omega$cm, $\rho_p = 3000\ \Omega$cm) were pre-melted in a single arc oven under argon atmosphere. Consecutively, single crystals were grown using the triarc Czochralski technique. The good crystallinity was evidenced by Laue analysis and by low energy electron diffraction (LEED) during photoemission experiments. \cite{zur07,klein08} A very low concentration of magnetic impurities in the pure FeSi crystals is derived from measurements of the magnetic susceptibility.

For the optical measurements, individual samples were obtained by cutting the single crystals into slices; the cut surfaces were  then polished to optical grade using a 0.25 $\mu$m diamond suspension. The samples were not oriented for these experiments which, however, does not affect the results of the optical investigations since the crystal structure is cubic.

\begin{figure}[b!]
\centering
\includegraphics[scale = 1]{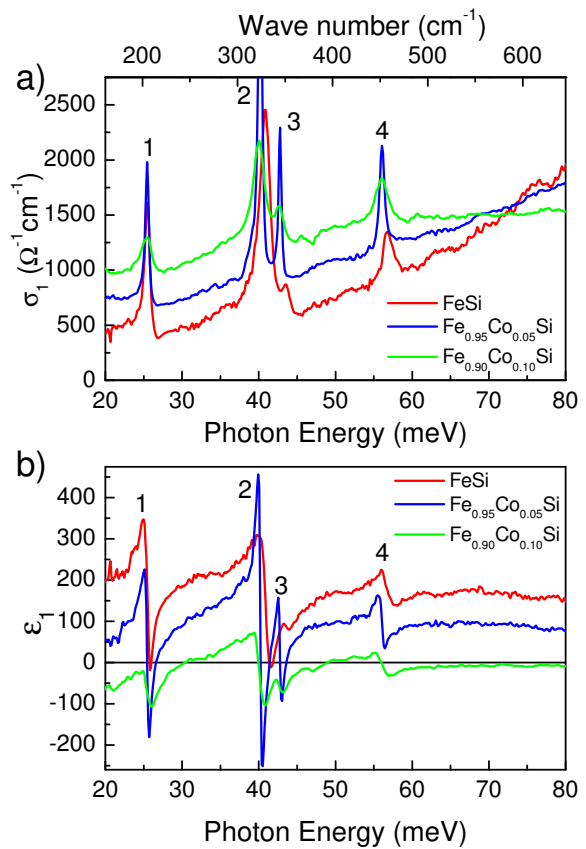}
\caption{(Color online) Real parts of (a) optical conductivity $\sigma_1(\omega)$    and (b) dielectric function $\varepsilon_1(\omega)$  of  Fe$_{1-x}$Co$_x$Si with various Co concentrations $x$ at $T=7$~K. Phonons are labeled from '1' to '4'.}
\label{fig1}
\end{figure}

The IR optical measurements were performed with home-built ellipsometers, attached to Fourier transform Bruker IFS-66vs spectrometers. As a source we used synchrotron edge radiation at the IR1 beamline of the ANKA synchrotron for the range of $10-85$~meV ($80-680$~cm$^{-1}$), and a SiC  glow bar and a mercury arc lamp for the range from 0.06 to 1.1~eV ($500-9000$~cm$^{-1}$). The angle of incidence was selected to 80.0$^\circ$ for all measurements. The infrared spectroscopic investigations were complemented with measurements in the range $0.8-6$~eV using a commercial Woolam VASE variable angle rotating-analyzer ellipsometer. In all these setups the samples were mounted in continuous helium flow cryostats ($6-450$~K). The base pressure at room temperature was  better than  5$\times$10$^{-7}$ mbar for the IR and 5$\times$10$^{-9}$ mbar for the visible setup.

The complex reflectance ratio $\rho$ was derived for an isotropic medium by inverting the Fresnel equations for the measured ellipsometric parameters $\Psi$ and $\Delta$, defined as
\begin{equation}
\tan \Psi e^{i \Delta}=\frac{r_p}{r_s}=\rho .
\label{Eq:elli}
\end{equation} 
Here $r_p$ and $r_s$ are the Fresnel reflection coefficients for light polarized parallel and perpendicular to the plane of incidence, respectively.  For a semi-infinite isotropic sample the complex dielectric function $\varepsilon(\omega)$ follows as

\begin{equation} \varepsilon(\omega)= \varepsilon_1(\omega) +i\varepsilon_2(\omega)=\sin^2\theta\left[ 1+\left( \frac{1-\rho}{1+\rho} \right)^{2}\tan^{2}\theta \right]\end{equation}

where $\varepsilon_1$, $\varepsilon_2$, and $\theta$ are the real and imaginary parts of the dielectric function and the angle of incidence, respectively.

\section{\label{sec:Result}Results and discussion}

\subsection{Dielectric response in the phonon energy range}

In our far-infrared spectra of FeSi at $T = 11$~K, four infrared active phonons are observed at 25.5~meV, 40.8~meV, 43.6~meV, and 56.8~meV (206 cm$^{-1}$, 329 cm$^{-1}$, 352 cm$^{-1}$ and 458 cm$^{-1}$, respectively) (Fig. \ref{fig1}), in good agreement with previous reports.\cite{schlesinger93,damascelli97,mena06} These modes are superimposed on a featureless electronic background. The optical conductivity, $\sigma_1(\omega)$, of the flat electronic background increases monotonically with the Co concentration, making the material more metallic  (Fig. \ref{fig1}a). This property is also reflected by the high negative values of $\varepsilon_1(\omega)$ at low photon energies (Fig. \ref{fig1}b).

\begin{figure}[h!]
  \centering
  \includegraphics[scale=1]{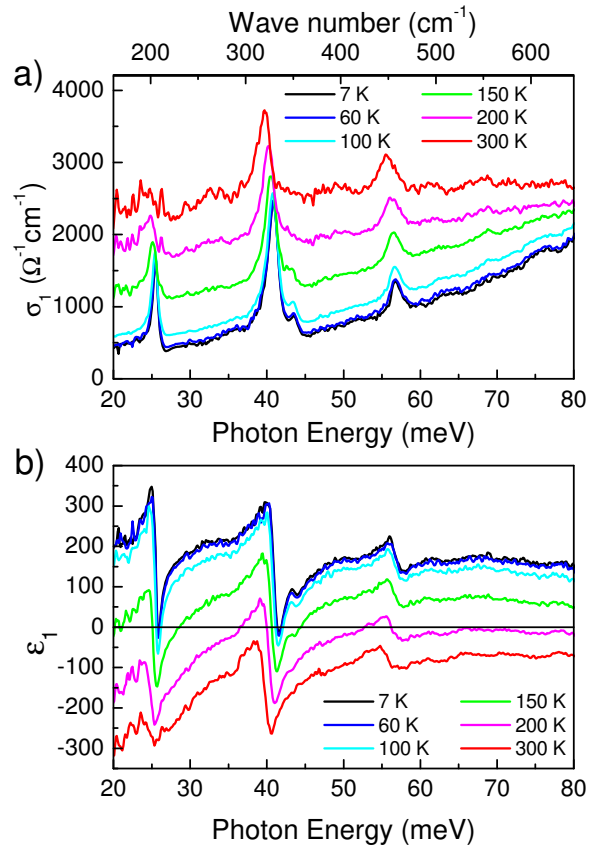}
  \caption{(Color online) Real parts of (a) optical conductivity $\sigma_1(\omega)$    and (b) dielectric function $\varepsilon_1(\omega)$  of FeSi measured at different temperatures. } 
  \label{fig2}
\end{figure} 

The influence of the temperature on the dielectric function of FeSi is displayed in Fig. \ref{fig2}. With increasing temperature, the gap is filled, leading to more metallic behavior. Therefore,  $\varepsilon_1(\omega)$ decreases and becomes negative as  $\sigma_1(\omega)$ rises, and the gap, which is observable at low temperatures, washes out. This result is in accordance with the observation made by Raman spectroscopy.\cite{racu07} From the evolution of the Raman phonon linewidth with temperature, it is shown that the chemical potential, which at zero temperature lies in the middle of the band gap, is shifted toward lower energy and crosses the top of the valence band near 250~K. 

In general, it is expected that in metals the phonons are screened by free electrons and, thus, should have a vanishing dipolar moment. However, even in the metallic regime of Fe$_{1-x}$Co$_x$Si, realized either by Co-doping or by increasing the temperature, the phonons are still observable in our infrared spectra.

A detailed analysis of the FeSi phonon response in the dielectric function shows that the inner phonons '2' and '3' (at 40.8 meV and 43.6 meV, respectively) are both fully symmetric and can be fitted by classical harmonic Lorentzian oscillators. In contrast, the two outermost phonons, '1' and '4' (at 25.5 meV and 56.8 meV, respectively), have asymmetric line shapes that cannot be described by the Lorentzian equation. In insulating systems having more than one infrared-active phonon mode at ${\bf k} = 0$ these modes are not necessarily independent, but can interact with each other. \cite{barker64} Therefore, the observed asymmetry suggests a coupling between the phonon modes. The dielectric function with $n$ phonons can be expressed by modified Lorentzian oscillators:\cite{humlicek00}

\begin{figure}[b!]
  \centering
    \includegraphics[scale=0.85]{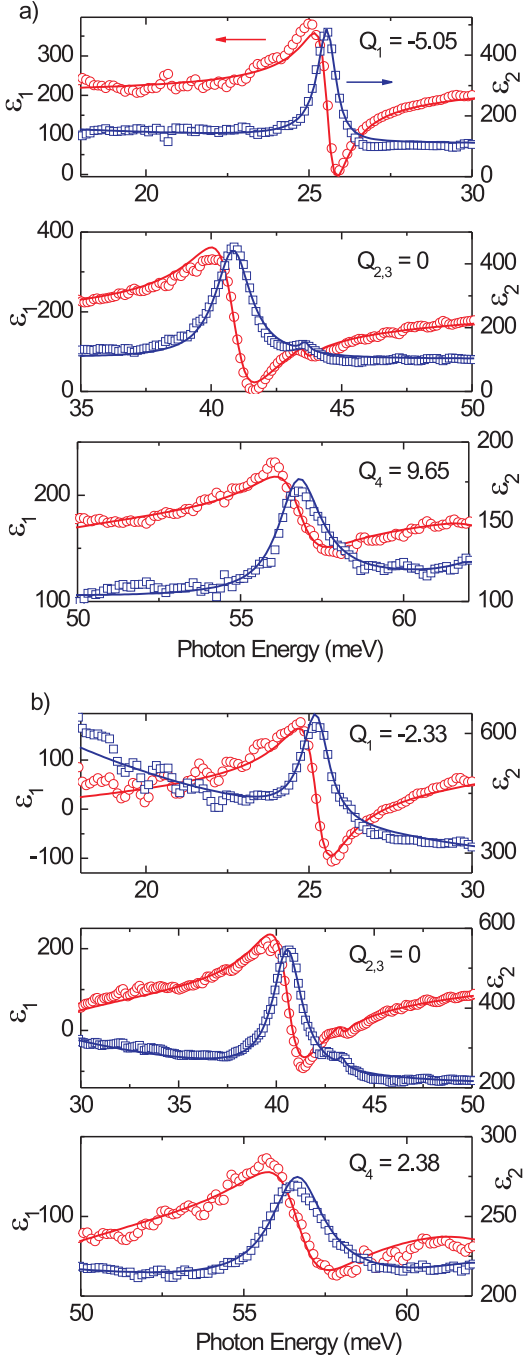}
  \caption{(Color online) Real  (open circles) and imaginary (open squares) parts of the measured complex dielectric response $\varepsilon(\omega)$ of FeSi. The solid curves present the result of the fitting   with Eqn. (\ref{humeqn})  at (a) 7~K and (b) 150~K ($Q_j$ values in meV).}
  \label{fig3}
\end{figure}

\begin{figure}[h!] 
  \centering
  \vspace{-10pt}
  \includegraphics[scale=0.4]{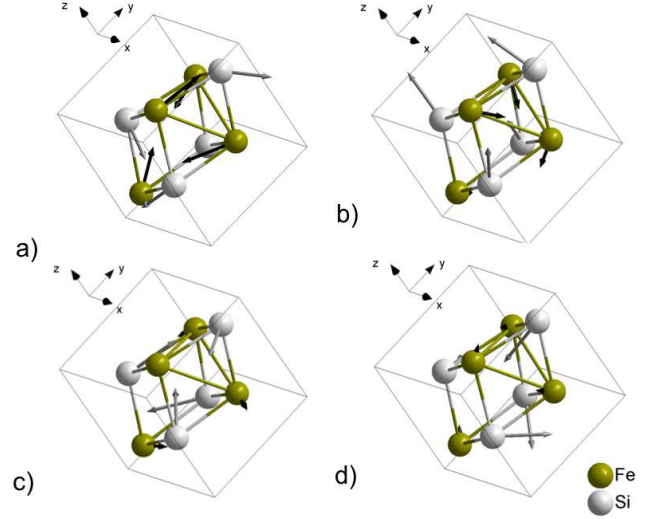}
  \caption{ (Color online) Eigenvectors for the observable IR-active modes at (a) 25.5 meV, (b) 40.8 meV, (c) 43.6 meV, and (d) 56.8 meV. Length of arrows represents the scaled magnitude of the vibrations.} 
  \label{fig4}
\end{figure}

\begin{equation}
\varepsilon (\omega) =\varepsilon_1 (\omega) + i \varepsilon_2 (\omega) = \varepsilon_\infty + \sum^n_{j=1} S_j \frac{\Omega^2_j -i \omega Q_j}{\Omega^2_j - \omega^2 -i \omega \Gamma_j}
\label{humeqn}
\end{equation}

where $\Omega_j$ and $\Gamma_j$ are the resonance frequency and the damping of the $j^{th}$ phonon, respectively, $S_j$ is the oscillator strength, and $Q_j$ (having the dimensions of frequency) denotes the coupling of the modes. For $Q_j = 0$, the $j$th phonon behaves like an independent classical oscillator. In absence of interaction with electronic states the sum of the phonon parameters $Q_j$ must vanish: 

\begin{equation}
\sum^n_{j=1} Q_j = 0
\label{sumrule}
\end{equation}

The four phonons observed can be successfully fitted by Eq.  (\ref{humeqn}) superimposed on a broad background (Fig.  \ref{fig3}). Our fit parameters for the phonons in FeSi at 7~K and 150~K are given in Table \ref{tab-D}. The asymmetry parameters $Q_{2,3}$ for phonons '2' and '3' are equal to zero, representing fully symmetric line-shapes. This implies that these two phonons are independent classical oscillators. Phonons '1' and '4' have finite parameters $Q_j\neq0$ with opposite sign, indicating a coupling between these modes. The sum rule in Eq.  (\ref{sumrule}) is fulfilled at 150~K (Fig. \ref{fig3}b), but not at 7~K (Fig. \ref{fig3}a). This behavior will be discussed further in Sec. \ref{el-ph}.

\begin{table}[b!]
\caption{Phonon fit parameters for of FeSi at 7~K (and 150~K).}
\label{tab-D}
\begin{tabular}{cccccc}
\hline \hline
 Phonon $j$ && $\Omega_j$ & $\Gamma_j$ & $S_j$ & $Q_j$\\
 & & (meV) & (meV) & &(meV) \\
\hline
'1' & & 25.5 (25.2) &  0.67 (0.99) & 9.40 (10.40) & -5.05 (-2.33)
 \\
\hline
'2' & & 40.8 (40.6) & 1.66 (1.80)  & 13.9 (13.7)& 0 (0)\\
\hline
 '3' & & 43.6 (43.3) & 0.39 (0.39)  &  0.25 (0.19) &  0 (0)\\
\hline
'4' & & 56.8 (56.5) & 1.61 (2.07)  & 1.93 (2.36) & 9.65 (2.38)\\
\hline \hline
\end{tabular}
\end{table}

The factor group analysis for FeSi yields 9 optical phonons, $\Gamma=2A + 2E + 5T$, which are all identified by Raman spectroscopy.\cite{racu07} The five T-modes are also IR-active.\cite{damascelli97} Density functional calculations employing the local density approximation have been performed with the electronic structure code {\sc CRYSTAL}06.\cite{CRYSTAL06} A Gaussian basis set is used, where the atom-centered basis functions are chosen as in Ref.  [\onlinecite{neef06}]. The initial step consists of a geometry optimization. In the next step, the second derivatives of the energy with respect to the atom positions are obtained using analytical first derivatives. Finally, the second derivatives are calculated numerically by displacing atoms. In this way, the mass-weighted Hessian matrix is obtained, and diagonalization gives the eigenvalues and eigenvectors.

Of the five computed IR-active $T$-modes, predicted at energies of 27.8~meV, 31.6~meV, 41.2~meV, 47.6, and 57.5~meV, four modes are observed in our IR-spectra at 25.5~meV, 40.8~meV, 43.6~meV, and 56.8~meV. The transition matrix element of the fifth phonon mode, which is expected at 31.6~meV,  is too small for the mode to be observable in the infrared spectra. The eigenvectors for the observed IR-active phonons are shown in Fig. \ref{fig4}. The 47.6~meV and 57.5~meV modes are mainly related to silicon motion.

\subsection{Band gaps of $\rm \bf Fe_{1-{\it x}}Co_{\it x}Si$}

Substituting Co for Fe in FeSi leads to a paramagnetic semiconductor for $x < 0.05$ and a ferromagnetic metal for $0.05 < x < 0.8$.\cite{manyala00} However, the size of the band gap in the semiconducting regime has not been reported consistently. From optical reflectivity measurements, a band gap of $\approx$80~meV has been reported. \cite{chernikov97} In contrast, photoemission data reveal a considerably smaller value for the gap of $\approx$30 meV.\cite{zur07} Here we resolve this apparent contradiction by giving evidence for the existence of both a direct ($\Delta E_{\rm dir}$) and an indirect band gap ($\Delta E_{\rm ind}$) in FeSi.

\begin{figure}[b!]
 \centering
 \includegraphics[scale=0.95]{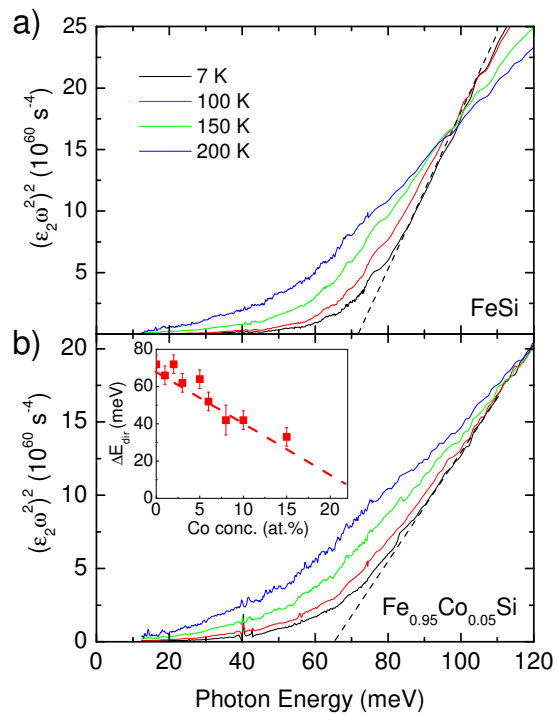}
 \caption{(Color online) Plot of  $(\varepsilon_{2}(\omega)\ \omega^{2})^2$ after subtraction of the phonon response at different temperatures for (a) FeSi and  (b) Fe$_{0.95}$Co$_{0.05}$Si  as a function of photon energy. The intercept of the dashed line with the energy axis defines the direct energy gap. Inset: Dependence of the direct gap of Fe$_{1-x}$Co$_x$Si on the Co concentration $x$. The dashed line is a guide to the eye.}
  \label{fig5}
\end{figure}

\textit{Direct gap.} In the case of a direct optical transition, the imaginary part of the dielectric function $\varepsilon_2(\omega)$ can be calculated as

\begin{equation}
\varepsilon_2 = A \frac{\sqrt{\frac{\hbar \omega}{\Delta E_{\rm dir}}-1}}{(\frac{\hbar \omega}{\Delta E_{\rm dir}})^2}
\end{equation}
with 
\begin{equation}
A = \frac{2e^2(2\mu)^{3/2}}{m_e^2\hbar}|P_{cv}|^2\Delta E_{\rm dir}^{-3/2}
\end{equation}

where $\mu =  \left( m_c^{-1} + m_v^{-1} \right)^{-1}$ is the reduced mass of the effective masses in the valence and the conduction band and $|P_{cv}|$ is the momentum matrix element. \cite{yucardona}
From these equations one can derive
\begin{equation}
(\varepsilon_2 \omega^2)^2 \propto \hbar \omega - \Delta E_{\rm dir}
\label{direct}
\end{equation}

\begin{figure}[t!]
  \centering
    \includegraphics[scale=0.8]{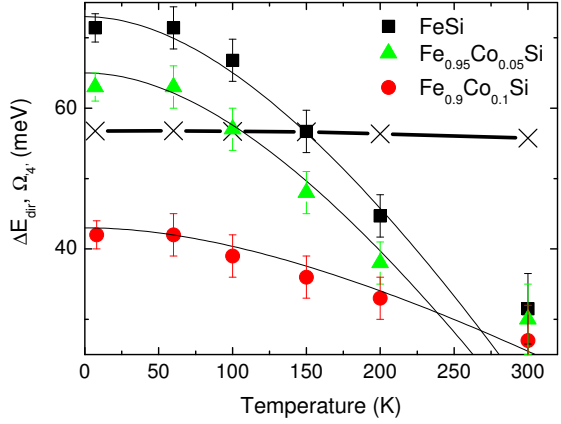}
  \caption{(Color online) Temperature dependence of the direct gap $\Delta E_{\rm dir}$ of Fe$_{1-x}$Co$_{x}$Si, for $x=0$, 0.05 and 0.1. The solid lines represent the fits using Eqn. (\ref{varshni}). The line with crosses ($-$$\times$$-$) denotes the frequency of phonon '4' in FeSi. Variation of $\Omega_{'4'}$  for different $x$ is smaller than 1 meV.}
  \label{fig6}
\end{figure}

\begin{table}[b!]
\caption{Parameters of the Varshni fit for 0\,\%, 5\,\% and 10\,\% Co-doped FeSi}
\label{param}
\begin{tabular}{cccc}\hline \hline
 Co conc.\quad\quad\quad\ & $E_{\rm dir.}(0)$ (meV)\quad\quad\quad\ &  $\alpha$\quad\quad\quad\ & $\beta$ (K) \\\hline 
0 at.\%\quad\quad\quad\ & 73\quad\quad\quad\ & 0.00047\quad\quad\quad\ & 490 \\
5 at.\%\quad\quad\quad\ & 63\quad\quad\quad\ & 0.00043\quad\quad\quad\ & 480 \\
10 at.\%\quad\quad\quad\ & 43\quad\quad\quad\ & 0.00015\quad\quad\quad\ & 470 \\\hline \hline
\end{tabular}
\end{table}

When $(\varepsilon_2 \omega^2)^2$ is plotted versus the photon energy, the direct band gap is obtained by the intersection of the fitted lines and the energy axis (e.g., for FeSi and  Fe$_{0.95}$Co$_{0.05}$Si in Fig. \ref{fig5}). For FeSi, a direct band gap of 73~meV at 7~K is found, which is in reasonable accordance with the results from earlier optical experiments.\cite{chernikov97} This direct gap still exists for 5 at.\% Co and does not even close for 15 at.\% (see inset of Fig.~\ref{fig5}b). As a function of temperature, the direct band gap follows the empirical Varshni equation \cite{varshni67} (Fig.~\ref{fig6}):

\begin{equation}
E_{\rm dir}(T) = E_{\rm dir}(0)-\frac{\alpha T^2}{T + \beta}
\label{varshni}
\end{equation}

The parameters obtained from the fit using Eq. (\ref{varshni}) are summarized in Tab. \ref{param}. 

\textit{Indirect gap.}   For parabolic bands, $\varepsilon_2$ follows from the equation \cite{made}
\begin{eqnarray}
\label{indirect}
\varepsilon_2 & = & C_{jj'}^{\rm abs} \omega^{-2} (\hbar \omega + \hbar \omega_{\rm ph} - \Delta E_{\rm ind}) +\\
\nonumber & + & C_{jj'}^{\rm em} \omega^{-2} (\hbar \omega - \hbar \omega_{\rm ph} - \Delta E_{\rm ind}) 
\end{eqnarray}
which includes the matrix elements $C_{jj'}$ for the absorption and the emission of a phonon. For indirect interband transitions, a plot of $(K\cdot \hbar \omega)^{1/2}$ vs. the photon energy (where $K$ is the absorption coefficient) may show linear intercepts that can be attributed to the absorption of charge carriers across the gap involving phonon emission and/or absorption, from which the size of the indirect gap can be derived.\cite{macfarlane} 

\begin{figure}[h!]
\begin{center}
\includegraphics[scale = 1]{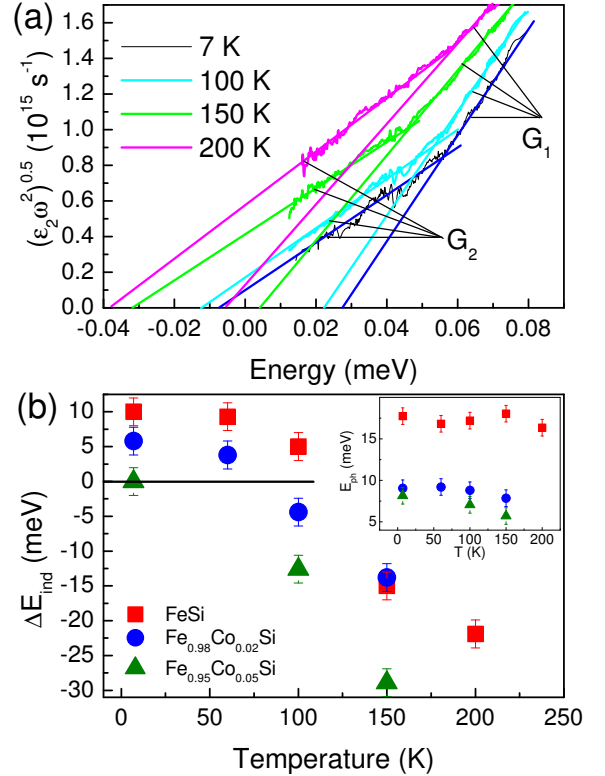}
\end{center}
\caption{(Color online) (a) Plot of $(\varepsilon_2 (\omega) \ \omega^2)^{1/2}$ of FeSi after subtraction of the phonon response at different temperatures. The straight lines represent the linear relation of  Eq. (\ref{eps_square}). (b) Temperature dependence of the indirect band gap energy. Inset: Energy of the transition-assisting phonon vs. temperature.}
\label{fig7}
\end{figure}

Because
\begin{equation}
\label{eps_square}
\varepsilon_2 = 2nk\ {\rm and}\ K \propto \frac{k}{\lambda} \propto k\omega\ \Rightarrow\ \varepsilon_2\omega^2 \propto K\omega,
\end{equation}
we plot $(\varepsilon_2\omega^2)^{1/2}$ as function of photon energy in Fig. \ref{fig7}. 
For analysis of the indirect gap, the modeled phonon modes have been subtracted from the imaginary part of the dielectric function, $\varepsilon_2(\omega)$. For each temperature, two linear intercepts (G$_1$ and G$_2$) may be attributed to phonon emission and absorption, respectively. However, the indirect absorption is temperature dependent due to the phonon assistance. The absorption of a phonon is proportional to the Bose distribution function $\left( e^{(E_{\rm ph}/k_BT)} -1 \right)^{-1}$, which has a value of only $\approx 10^{-9}$ for $T = 7$~K and a phonon with an expected energy in the order of 10~meV. Therefore, at low temperatures below 200 K, only absorption branches due to the emission of a phonon can be observed, and these can only take place at photon energies $\hbar \omega > \Delta E_{\rm ind} + E_{\rm ph}$. For $\Delta E_{\rm ind} \approx 30$~meV, as derived in our previous photoemission\cite{menzel07,zur07} and Raman experiments,\cite{racu07} and $E_{\rm ph} \approx 10$~meV, $\hbar\omega$  should exceed 40~meV. Thus, only the linear slopes $G_1$ can be attributed to an indirect transition with the emission of a phonon; the lines $G_2$ cannot be assigned to interband transitions, which are assisted by thermally excited phonons.

Because the ellipsometric measurements were performed using a Fourier spectrometer, the sample was illuminated by synchrotron radiation in the spectral range between 0.4~meV and 1.2~eV. Even if it is assumed that only 1\% of the synchrotron beam flux reaches the sample, the surface is irradiated with $\approx 10^{11}$ photons per second. This leads to a continuous generation of optical phonons, which may also decay into low-energy acoustic phonons. Therefore, a sufficient number of optically excited phonons may allow for phonon absorption, even at low temperatures. If so, one is able to derive the absolute value of the indirect gap from the arithmetic mean of the intersects of $G_1$ and $G_2$ with the energy axis. At 7~K, the indirect gap for FeSi is 10~meV and decreases with increasing temperature (Fig. \ref{fig7}b). The indirect gap changes sign and becomes negative above 100~K. This manifests as a crossover from semiconducting to semimetallic behavior, where the valence and conductivity bands approach each other and overlap at different points of the Brillouin zone.

The indirect gap is also reduced in Fe$_{1-x}$Co$_x$Si with increasing Co concentration and closes at low temperature in Fe$_{0.95}$Co$_{0.05}$Si. The phonon that assists the indirect transition in FeSi has an energy of $\hbar\omega_{\rm ph} \approx 17$~meV, which is temperature independent within the accuracy of the measurement. The phonon energy is smaller than that of all optical phonons known from factor group analysis and Raman spectroscopy,\cite{racu07} which therefore implies that an acoustic phonon assists the optical transition.

\begin{figure}[b!]
\begin{center}
\includegraphics[scale = 0.43]{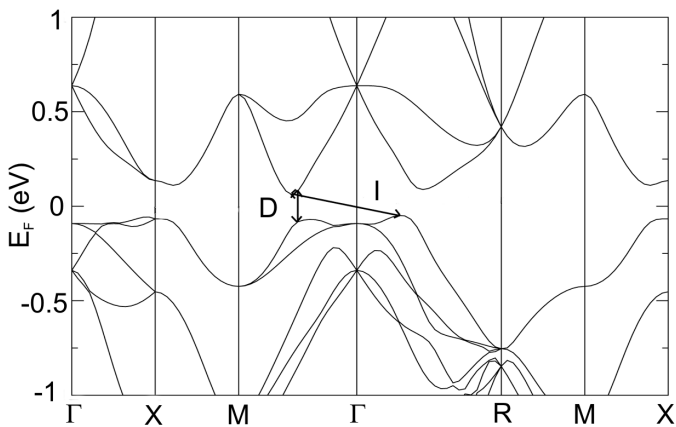}
\end{center}
\caption{Calculated band structure of FeSi using the density functional theory. Both indirect (I) and direct (D) transitions are shown in the band structure.}
\label{fig8}
\end{figure}

 In the band structure calculation, using density functional theory with the gradient corrected functional of Perdew and Wang (PW91), three indirect transitions between $\Gamma$--$M$, $\Gamma$--$R$, and $\Gamma$--$X$ can be identified (Fig. \ref{fig8}). The minimum direct gap (0.18~eV) occurs along the $\Gamma$--$M$ direction and the minimum indirect gap corresponds to transitions from the valence band in the $\Gamma$--$R$ direction to the conduction band in the $\Gamma$--$M$ direction (0.12~eV). It is to be noted here that the calculated energy values are overestimated compared with the experimentally measured values. This is evidenced in high-resolution photoemission experiments, which reveal a renormalization of the Fe bands close to the Fermi energy due to the self-energy of the Fe-3$d$ electrons.\cite{klein08} Because the transition matrix elements are not accessible from this calculation, it is not clear which of the three possible indirect transitions is dominant in the infrared spectrum. The minimum gap energies are indicated in Fig. \ref{fig8}. 
The gap observed in photoemission (PE) experiments \cite{zur07} (28~meV) is slightly larger than the indirect gap derived from the ellipsometric measurements. However, because the data in the PE spectra are restricted to a limited section of the Brillouin zone, the gap observed in a certain crystalline direction is not necessarily the smallest one. 

Contrary to the suggestions in Refs.  \onlinecite{schlesinger93} and \onlinecite{damascelli97}, the closing of the gap in FeSi at 200 K seems to be due to indirect transitions, rather than to the direct gap that is not expected to close until 700~K; from this point of view, there is no need to infer strongly correlated electrons in this compound.

\subsection{Electron-phonon interaction} \label{el-ph}

As we have discussed in Section IIIA the inner phonons are classical oscillators, whereas phonons '1' and '4' show strong anharmonicity. The origin of  the phonon anharmonicity is due to phonon-phonon and electron-phonon interactions. The parameter $Q$ is introduced in  Eq.~(\ref{humeqn}) to describe the asymmetric lineshape  of the anharmonic phonons. For insulating materials with strong phonon-phonon interaction in absence of electron-phonon interaction, the Eq.~(\ref{sumrule}) is satisfied. We found that this is not the case in FeSi, where phonon '1' has a much less asymmetric lineshape than phonon '4', $\left\vert  Q_1\right\vert  < \left\vert Q_4\right\vert$. This indicates that the additional lineshape distortion of  phonon '4' comes from the interaction of this phonon with electronic states, forming the absorption edge.

\begin{figure}[b!] 
  \centering
  \includegraphics[scale=1]{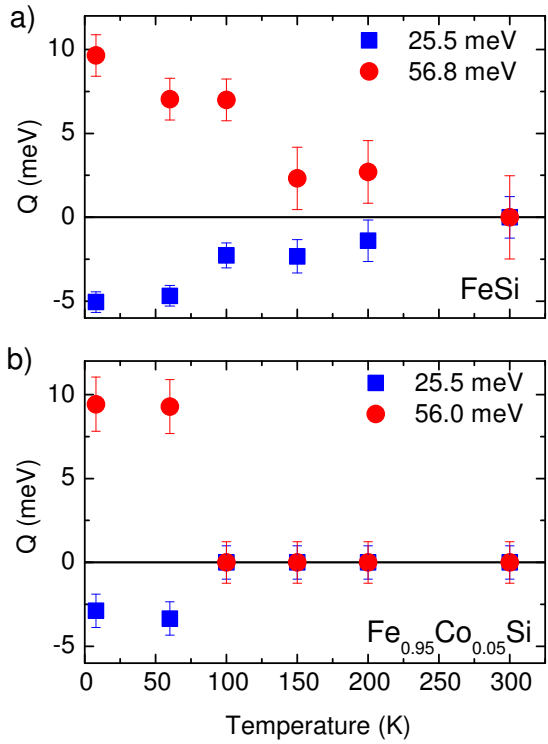}
      \caption{(Color online) Temperature dependence of the asymmetry parameter $Q$ for phonons '1' and '4' in (a) FeSi and (b) Fe$_{0.95}$Co$_{0.05}$Si.}
  \label{fig9}
\end{figure}

The shape of the broad electronic continuum in the optical conductivity in Fig.~\ref{fig2}a has a steep positive slope near the absorption edge. Because phonon '4'  is energetically close to the direct absorption edge
one would expect that the damping of the longitudinal LO  vibration is much higher than that of the corresponding transverse  TO vibration, causing the strong lineshape asymmetry.
Moreover, both types of damping (LO and TO) have different temperature dependencies, because the electronic background is getting flatter with increasing temperature; both parameters $Q_1$ and $Q_4$ and their sum tends to zero. The direct absorption edge shifts towards lower energies and crosses the energy of the fourth TO phonon at $\approx$150~K in FeSi and $\approx$100~K in Fe$_{0.95}$Co$_{0.05}$Si (Fig. \ref{fig6}). Above  these temperatures  the damping of the LO and TO vibrations for phonon '4' becomes equal  (Fig. \ref{fig9}).
Additionally, the indirect gap is filled at higher temperatures and the charge carriers screen out the phonon-phonon interaction. Therefore a symmetric lineshape for all four phonons is observed.

 Increasing the Co-doping enlarges the electronic background, causing the outermost phonons become more and more symmetric. In case of higher doping (more than 10\% Co), the phonon asymmetry vanishes as the enhanced electronic background screens out the coupling between the phonons.

\begin{figure}[h!]
  \centering
    \includegraphics[scale=1]{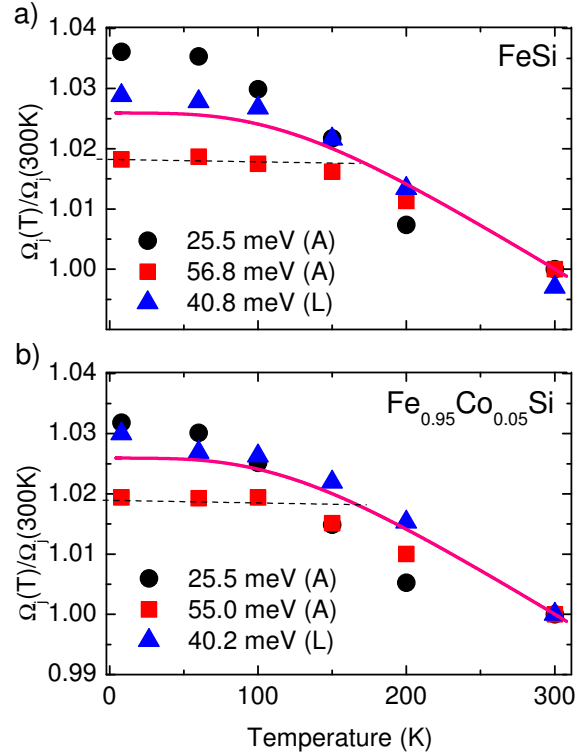}
    \caption{(Color online) Relative changes of phonon frequencies $\Omega_J(T)$ for the two asymmetric (A) and one Lorentzian (L) modes in (a) FeSi and (b) Fe$_{0.95}$Co$_{0.05}$Si  with decreasing temperature below 300~K. The solid lines represent the expected frequency change due to thermal expansion. The dashed lines are guides to the eyes (see text).}
  \label{fig10}
\end{figure}

There are two contributions to the change of phonon frequency with temperature: the shift arising from the volume change due to thermal expansion and the phonon-phonon interaction. The phonon frequency shift, $\delta\omega$, due to anharmonicity effects is given as: \cite{sakurai71}

\begin{equation}\delta \omega =\omega_0 \left(  e^{-\gamma\int_0^T\alpha_V dT} -1\right) ,  \end{equation}

where $\omega_0$, $\gamma$, $\alpha_V$ are the phonon frequency at zero temperature, the mode Gr\"uneisen parameter, and the volumetric thermal expansion coefficient, respectively. We used one Gr\"uneisen parameter for all modes, calculated by the relation $\gamma =\alpha_V V_{\rm molar}/(C_p\chi_S)$, taking the molar volume $V_{\rm molar}=13.6$~cm$^3$/mol.\cite{wood} The values of the specific heat, $C_p$, and the adiabatic compressibility, $\chi_S$, of FeSi are taken from [\onlinecite{takahashi00}] and [\onlinecite{mandrus95}], respectively. Using the thermal expansion values\cite{meingast08} for our FeSi samples and neglecting the phonon-phonon interaction, we estimate the Gr\"uneisen parameter at 3.4 for temperatures up to 300~K. The expected frequency changes due to the temperature-driven lattice expansion for FeSi and Fe$_{0.95}$Co$_{0.05}$Si are calculated using this Gr\"uneisen parameter and the thermal expansion data and are shown in Fig. \ref{fig10}.

  The fact that phonon '4' is coupled more strongly to the electronic states can be seen in Fig.\ref{fig10}. Phonons '1' to '3' follow the expected temperature dependence due to anharmonicity effects, whereas an effective softening can be observed for phonon '4'. This effective softening vanishes when the absorption edge crosses the phonon energy (Fig. \ref{fig6}).
  
  Despite the existence of a strong electron-phonon interaction, in the metallic regime the IR-active phonons are not fully screened by the conduction electrons. Because an electric dipole moment is necessary to excite a phonon optically, further insight into this peculiarity can be obtained by an analysis of the charge transfer between Fe and Si.

The generalized multimode Lyddane-Sachs-Teller relation \cite{lyddane59,kurosawa61} 
\begin{equation}
\prod_{i=1}^4 \frac{\omega_{\rm LO,i}^2}{\omega_{\rm TO,i}^2} = \frac{\varepsilon_{\rm stat}}{\varepsilon_{\rm opt}}
\end{equation}

which compares the values of the dielectric function with the phonon frequencies, is fulfilled to within 20\% and therefore represents a self-consistency check of our measurements (see Tab. \ref{tab-1}). With the derived transverse and longitudinal phonon frequencies, the Born effective charge $e_i^*$ can be expressed by \cite{kurosawa61}

\begin{equation}
N \sum_i \frac{e_i^*}{M_i} = \varepsilon_{\rm opt} \sum_j \left( \omega_{\rm LO,j}^2 - \omega_{\rm TO,j}^2 \right)
\end{equation}

\begin{table}[b!]
\caption{Self-consistency check using the generalized Lyddane-Sachs-Teller relation} 
\label{tab-1}
\begin{tabular}{cccc}
\hline \hline
 & & $\prod_{i=1}^4 \omega_{\rm LO,i}^2 / \omega_{\rm TO,i}^2$ & $\varepsilon_{\rm stat} / \varepsilon_{\rm opt}$ \\
\hline
 FeSi & & 1.03 & 1.22\\
\hline
 Fe$_{0.98}$Co$_{0.02}$Si & & 1.01 & 1.15\\
\hline
 Fe$_{0.95}$Co$_{0.05}$Si & & 1.03 & 1.23\\
\hline \hline \\
\end{tabular}
\end{table}

yielding $e_i^* = 2.4 \pm 0.6$ for all investigated Co concentrations $0 \leq x \leq 0.05$ (Tab. \ref{tab-2}). The consideration of polarization effects results in the Szigeti effective charge $e_s^*$, which includes the screening of the phonons due to free carriers\cite{szigeti50} by the expression of Clausius-Mosotti
\begin{equation}
e_s^* = e_i^* \frac{3}{\varepsilon_{\rm opt} + 2}.
\end{equation}

The Szigeti effective charge is lower than 0.1$e$ for all Co concentrations. Despite the nearly complete screening, the phonon response is still visible in the infrared spectra. From the density functional calculations, it is known that the additional electron charge from each Co dopand essentially resides on the Co site.\cite{racu07} Therefore, the oscillations between Fe and Si may not be affected by the charge density localized on the Co and, in addition, the ratio of Co with respect to Fe is small. A similar ineffective screening of the phonons is observed in the semiconductor U$_2$Ru$_2$Sn, where the low carrier concentration of $3.5\times 10^{20}$~cm$^{-3}$ is  responsible for the ineffective screening. \cite{sichelschmidt07} In FeSi the carrier concentration is on the order of several times $10^{18}$~cm$^{-3}$ at low temperature, even lower than in U$_2$Ru$_2$Sn.

In conclusion, two coupling effects can be observed. First, a coupling between phonons '1' and '4' exists as indicated by an asymmetry parameter of opposite sign and, second, the phonons are affected by an electron-phonon coupling that changes as a function of temperature because electronic decay channels appear at lower excitation energies when the gap is filled up. This result is corroborated by Raman experiments, in which the electron-phonon coupling is observed in terms of the broadening of the phonon line-shapes. The linewidth increases as function of temperature, and the broadening saturates at about 250~K as a consequence of the chemical potential crossing the top of the valence band. \cite{racu07}

\begin{table}[h!]
\caption{Born and Szigeti effective charge of Fe$_{1-x}$Co$_x$Si.}
\label{tab-2}
\begin{tabular}{cccc}
\hline \hline
 & & Born effective\quad\quad\quad & Szigeti effective \\
 & & charge e$^*_i$\quad\quad\quad & charge e$^*_S$\\
\hline
 FeSi & & 1.81\quad\quad\quad & 0.03 \\
\hline
 Fe$_{0.98}$Co$_{0.02}$Si & & 2.46\quad\quad\quad & 0.05 \\
\hline
 Fe$_{0.95}$Co$_{0.05}$Si & & 3.00\quad\quad\quad & 0.08 \\
\hline \hline
\end{tabular}
\end{table}

\subsection{High-energy spectral weight transfer}

In a conventional semiconductor, thermally excited charge carriers increase the far-IR conductivity at high temperatures. On the other hand, because the electrons partially occupy the conduction band and holes are present at the top of the valence band, absorption at energies just above the gap is reduced. For a conventional semiconductor, it is expected that the spectral weight lost below the gap is recovered, with increasing temperature, in an energy range on the order of a few times the gap. The integrated optical conductivity can be quantitatively analyzed by defining the restricted sum rule function:

\begin{equation}
N_{\rm eff}( \omega) = \frac{2mV}{\pi e^2 }\int^{\omega}_{0} \sigma_1(\omega) d\omega,
\end{equation}
where $N_{\rm eff}$, $m$, $e$, and $V$ are the effective number of electrons, the free-electron mass and charge, and the volume of one FeSi formula unit. In the limit $\omega\rightarrow \infty$ the effective number of electrons is equal to the total number of electrons in one formula unit. The lattice parameter for FeSi was taken from [\onlinecite{wood}] and used to calculate $N_{\rm eff}$ for all Co-doping. The estimated error of $N_{\rm eff}$ due to thermal expansion and doping is less than 1\,\%.

\begin{figure*}[t!]
   \includegraphics[scale=1.4]{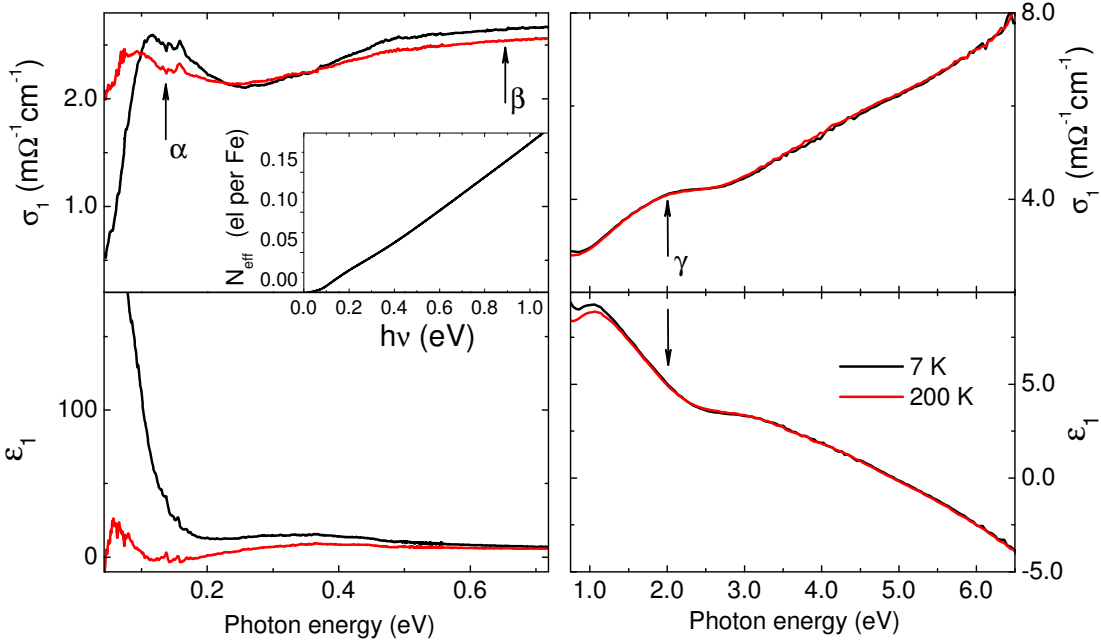}
\vspace{-30pt}
  \caption{(Color online) Real parts of  optical conductivity $\sigma_1(\omega)$    and  dielectric function $\varepsilon_1(\omega)$  for FeSi  at 7~K (black) and 200~K (red). Phonons are subtracted. Inset: Normalized effective charge density at 7~K. } 
  \label{fig11}
\end{figure*}

Figure~\ref{fig11} shows the real parts of the optical conductivity, $\sigma_1(\omega)$, and of the dielectric function, $\varepsilon_1(\omega)$, of FeSi in the range 40 meV to 6.5 eV for 7 K and 200 K (fitted phonons are subtracted). The left panels of Fig.~\ref{fig11} show that the optical response is strongly temperature dependent within several gap energies, $\Delta E_{\rm dir}\approx$ 73 meV. Above 1~eV, the optical conductivity is nearly temperature independent. From $N_{\rm eff}(\omega)$ integrated up to 1 eV (Fig.\ref{fig11}, insert), we estimate that 0.2 electrons per Fe atom participate in the optical transitions up to this energy.

\begin{figure}[b!] 
   \includegraphics[scale=0.8]{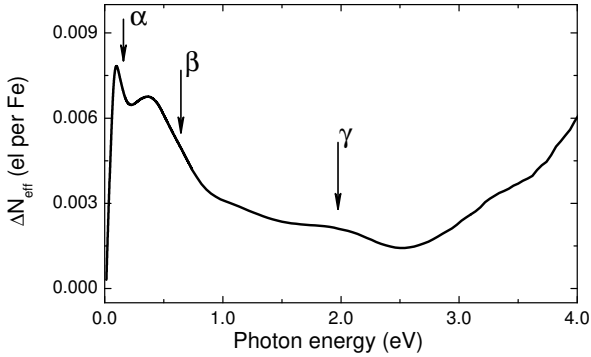}
   \caption{Effective charge density difference $\Delta N_{\rm eff}(\omega)$ between 200~K and 7~K in FeSi. Absorption bands at 0.15, 0.65 and 2.0 eV are shown with arrows.}
\label{fig12}
\end{figure}

In order to analyze the temperature dependent redistribution of the spectral weight, in Fig. \ref{fig12} we plot the difference between the normalized effective charge densities at 200~K and 7~K for FeSi, calculated as

\begin{equation}
\Delta N_{\rm eff} (\omega)= N_{\rm eff}^{200\;{\rm K}}(\omega) - N_{\rm eff}^{7\;{\rm K}}(\omega). 
\label{delta}
\end{equation}

A full recovery of the spectral weight happens when $\Delta N_{\rm eff}(\omega)$ approaches zero. In FeSi, $\Delta N_{\rm eff}(\omega)$ increases within the gap energy range below 0.1~eV, then decreases up to 2.5 eV, with a small bump at 0.35~eV, and remains positive throughout. One can identify three distinct absorption bands at 0.15~eV ($\alpha$), 0.65~eV ($\beta$), and 2.0~eV ($\gamma$), whose spectral weight is transferred to the far-IR range due to closing of the band gap. The following increase of $\Delta N_{\rm eff}$ above 2.5~eV can be attributed to the broadening of a strong transition above 4~eV with increasing temperature. This prevents the complete recovery of the spectral weight near 3~eV (Fig. \ref{fig12}), which is almost 40 times the direct optical gap energy.

The optical spectral weight transfer due to dynamic renormalization effects is widely seen in strongly correlated electron materials, which reflects the redistribution of the spectral weight between Hubbard bands. In this scenario, the optical spectral weight is recovered on the energy scale of on-site Coulomb interactions. Supporting this scenario, it has been reported that in FeSi, the spectral weight is not fully recovered up to at least 6~eV.\cite{schlesinger93,mena06}

Contrary to the previous reports, we find that the spectral weight is recovered within a 3~eV spectral range. Moreover, we find that FeSi is as an indirect gap semiconductor, exhibiting the temperature-driven semiconductor to semimetal crossover. In semimetallic materials, significant temperature-dependent changes in the optical spectrum are expected over a broad spectral range. This behavior arises primarily from changes in the valence and conduction band occupancies at different points of the Brillouin zone due to phonon-assisted indirect transitions.
Phonons are not available for vibronic processes at low temperature. As temperature increases, the phonon assistance is dramatically enhanced, both by the thermal excitation of phonons and by concomitant complete closing of the indirect gap. Our findings imply that the electron-phonon interaction and semimetallic character of FeSi play the dominant role in the observed broad-band spectral weight transfer, although one should do not completely exclude the relevance of the alternative scenario.

\section{Conclusion}

We have presented spectroscopic ellipsometry studies of Fe$_{1-x}$Co$_x$Si ($x=0-0.2$) single crystals. Our optical phonon spectra show four IR-active modes, in agreement with previous reports. We suggest the assignment of all phonon modes observed by comparing them with the results of density functional calculations. A detailed analysis of the phonon response in the dielectric function of FeSi shows that the 'inner' phonons at 40.8~meV (329 cm$^{-1}$) and 43.6~meV (352 cm$^{-1}$) are fully symmetric, whereas the 'outermost' phonons at 25.5~meV (206 cm$^{-1}$) and 56.8~meV (458 cm$^{-1}$) have asymmetric line shapes, which cannot be described by the Lorentzian equation. The asymmetry parameters introduced for these modes have opposite sign, indicating a direct coupling between them. Moreover, the phonons are affected by an electron-phonon coupling that changes with temperature, evidenced by electronic decay channels that appear at lower excitation energies when the band gap is filled up. The electron-phonon coupling is considered a possible origin of phonon anharmonicity. When the absorption edge shifts across the phonon energy, following the temperature dependence of the direct band gap, the phonon line-shape asymmetry vanishes. The self-consistency of the phonon spectra is verified using the generalized Lyddane-Sachs-Teller relation. The Born e$^*_i$ and Szigeti e$^*_S$ effective charges are calculated to determine the charge transfer between Fe and Si, yielding e$^*_i = 2.4 ± 0\pm6$ and e$^*_S < 0.1$ for all investigated Co concentrations $0 \leqslant x \leqslant 0.05$.

Direct and indirect band gaps of 73~meV and 10~meV, respectively, are observed in FeSi at 7~K. Systematic investigations of the nature of the gaps in the Fe$_{1-x}$Co$_x$Si system, together with its temperature and doping dependence, have been carried out. The temperature dependence of the direct gap is well described by the semiempirical Varshni equation. As temperature increases, the indirect gap changes sign, which manifests as a crossover from semiconducting to semimetallic behavior.

Finally, we address the question of what causes the enhancement in the far-IR spectral weight due to closing of the optical gaps. We identified three distinct absorption bands at 0.15~eV, 0.65~eV and 2.0~eV whose spectral weight is transferred to the far-IR range with increasing temperature. Our findings imply that the electron-phonon interaction and semimetallic character of FeSi play a dominant role in the broad-band spectral weight transfer. We conclude that the optical properties of Fe$_{1-x}$Co$_x$Si can be well described in an itinerant picture, taking self-energy corrections into account.

\section{Acknowledgments}
We wish to acknowledge the support of Y.-L. Mathis at the IR 1 beamline of ANKA. We are grateful to Q.~Zhang and C.~Meingast for providing the thermal expansion data prior to publication.


\begin{thebibliography}{99}
\bibitem{aeppli92} G. Aeppli and Z. Fisk, Comments Cond. Mat. Phys. {\bf 16}, 155 (1992).
\bibitem{jaccarino67} V. Jaccarino, G. K. Wertheim, J. H. Wernick, L. R. Walker, and S. Arajs, Phys. Rev. {\bf 160}, 476 (1967).
\bibitem{mandrus95} D. Mandrus, J. L. Sarrao, A. Migliori, J. D. Thompson, and Z. Fisk, Phys. Rev. B {\bf 51}, 4763 (1995).
\bibitem{mattheiss93} L. Mattheiss and D. R. Hamann, Phys. Rev. B {\bf 47}, 13114 (1993).
\bibitem{chernikov97} M. A. Chernikov, L. Degiorgi, E. Felder, S. Paschen, A. D. Bianchi, H. R. Ott, J. L. Sarrao, Z. Fisk, and D. Mandrus, Phys. Rev. B {\bf 56}, 1366 (1997).
\bibitem{nyhus95} P. Nyhus, S. L. Cooper, and Z. Fisk, Phys. Rev. B {\bf 51}, R15626 (1995).
\bibitem{zur07} D. Zur, D. Menzel, I. Jursic, J. Schoenes, L. Patthey, M. Neef, K. Doll, and G. Zwicknagl, Phys. Rev. B {\bf 75}, 165103 (2007).
\bibitem{klein08} M. Klein, F. Reinert, D. Zur, D. Menzel, J. Schoenes, K. Doll, and J. R\"oder, Phys. Rev. Lett. {\bf 101}, 046406 (2008).
\bibitem{neef06} M. Neef, K. Doll, and G. Zwicknagl, J. Phys.: Condens. Matter {\bf 18}, 7437 (2006).
\bibitem{menzel06} D. Menzel, M. Finke, T. Donig, A. M. Racu, and J. Schoenes, Physica B {\bf 378-380}, 718 (2006).
\bibitem{racu07} A. M. Racu, D. Menzel, J. Schoenes, and K. Doll, Phys. Rev. B {\bf 76}, 115103 (2007).
\bibitem{menzel04} D. Menzel, D. Zur, and J. Schoenes, J. Magn. Magn. Mater. {\bf 272-276}, 130 (2004).
\bibitem{wernick72} J. H. Wernick, G. K. Wertheim, and R. C. Sherwood, Mater. Res. Bull. {\bf 7}, 1431 (1972).
\bibitem{breuer97} K. Breuer, S. Messerli, D. Purdie, M. Garnier, M. Hengsberger, Y. Baer, and M. Mihalik, Phys. Rev. B {\bf 56}, R7061 (1997).
\bibitem{schlesinger93} Z. Schlesinger, Z. Fisk, H.-T. Zhang, M. B. Maple, J. F. DiTusa, and G. Aeppli, Phys. Rev. Lett. {\bf 71}, 1748 (1993).
\bibitem{degiorgi94} L. Degiorgi, M. B. Hunt, H. R. Ott, M. Dressel, B. J. Fenestra, G. Gr\"uner, Z. Fisk, and P. Canfield, Europhys. Lett. {\bf 28}, 341 (1994).
\bibitem{damascelli97} A. Damascelli, K. Schulte, D. van der Marel, and A. A. Menovsky, Phys. Rev. B {\bf 55}, R4863 (1997).
\bibitem{mena06}  F.P. Mena,  J.F. DiTusa, D. van der Marel, G. Aeppli, D.P. Young, A. Damascelli, and J.A. Mydosh, Phys. Rev. B {\bf 73}, 085205 (2006).
\bibitem{webb86} B. C. Webb, A. J. Sievers, and T. Mihalisin, Phys. Rev. Lett. {\bf 57}, 1951 (1986).
\bibitem{barker64} A. S. Barker and J. J. Hopfield, Phys. Rev. {\bf 135}, A1732 (1964).
\bibitem{humlicek00} J. Humli\v{c}ek, R. Henn, and M. Cardona, Phys. Rev. B {\bf 61}, 14554 (2000).
\bibitem{CRYSTAL06} R. Dovesi, V. R. Saunders, C. Roetti, R. Orlando, C. M. Zicovich-Wilson, F. Pascale, B. Civalleri, K. Doll, N. M. Harrison, I. J. Bush, Ph. D'Arco, M. Llunell,{\sc crystal 2006} User's Manual, University of Torino, Torino (2006).
\bibitem{manyala00} N. Manyala, Y. Sidis, J. F. DiTusa, G. Aeppli, D. P. Young, and Z. Fisk, Nature (London) {\bf 404}, 581 (2000).
\bibitem{made} O. Madelung, in: \textit{Festk\"orpertheorie}, Springer Berlin Heidelberg (1972).
\bibitem{yucardona} P. Y. Yu and M. Cardona in: \textit{Fundamentals of Semiconductors}, Springer Berlin Heidelberg (1996).
\bibitem{varshni67} Y. P. Varshni, Physica {\bf 34}, 149 (1967).
\bibitem{macfarlane} G. G. Macfarlane, T. P. McLean, J. E. Quarrington, and V. Roberts, Phys. Rev. {\bf 111}, 1245 (1958).
\bibitem{menzel07} D. Menzel, D. Zur, I. Jursic, J. Schoenes, L. Patthey, M. Neef, and L. Doll, J. Magn. Magn. Mater. {\bf 310}, 368 (2007).
\bibitem{man84} A. Manoogian and J. C. Woolley, Can. J. Phys. {\bf 62}, 285 (1984).
\bibitem{paschen97} S. Paschen, E. Felder, M. A. Chernikov, L. Degiorgi, H. Schwer, H. R. Ott, D. P. Young, J. L. Sarrao, and Z. Fisk, Phys. Rev. B {\bf 56}, 12916 (1997).
\bibitem{takahashi00} Y. Takahashi, T. Kanomata, R. Note, and T. Nakagawa, J. Phys. Soc. Jpn. {\bf 69}, 4018 (2000).
\bibitem{sakurai71} T. Sakurai and T. Sato, Phys. Rev. B {\bf 4}, 583 (1971).
\bibitem{wood} I. G. Wood, W. I. F. David, S. Hull, and G. D. Price, J. Appl. Cryst. {\bf 29}, 215 (1996).
\bibitem{meingast08} Q. Zhang and C. Meingast, unpublished
\bibitem{lyddane59} R. H. Lyddane, R. G. Sachs, and E. Teller, Phys. Rev. {\bf 59}, 673 (1959).
\bibitem{kurosawa61} T. Kurosawa, J. Phys. Soc. Jpn. {\bf 16}, 1298 (1961).
\bibitem{szigeti50} B. Szigeti, Proc. Roy. Soc. London A {\bf 204}, 51 (1950).
\bibitem{sichelschmidt07} J. Sichelschmidt, V. Voevodin, J. A. Mydosh, and F. Steglich, J. Magn. Magn. Mater. {\bf 310}, 434 (2007).
\end{thebibliography}
\end{document}